# Faceting Oscillations in Nano-Ferroelectrics

J. F. Scott[1] and Ashok Kumar[2]

[1]Cavendish Lab., Cambridge Univ., Cambridge, UK
[2]CSIR-National Physical Laboratory, Delhi, India

**Abstract:**

We observe periodic faceting of 8-nm diameter ferroelectric disks on a 10 s time-scale when thin $Pb(Zr_{0.52}Ti_{0.48})O_3$ (PZT) film is exposed to constant high-resolution transmission electron microscopy (HRTEM) beams. The oscillation is between circular disk geometry and sharply faceted hexagons. The behavior is analogous to that of spin structure and magnetic domain wall velocity oscillations in permalloy [A. Bisig et al., Nature Commun. 4, 2328 (2013)], involving overshoot and de-pinning from defects [C. P. Amann, et al., J. Rheol. 57, 149-175 (2013)].

Corresponding Authors:
Prof. J F Scott (jfs32@cam.ac.uk);
Dr. Ashok Kumar (ashok553@nplindia.org)

Historically ferroelectric domains have been treated in analogy with magnetic domains. Although the two have some superficial similarities with regard to static structure, their dynamics is fundamentally different in two important ways: First, the temporal dependence of spin waves is described by the Landau-Lifshitz-Gilbert Equations, and these equations are first-order in time. That requires that when the external magnetic field (H) stops, the spin precession stops instantly. By comparison, ferroelectric polarizations and domain walls obey Newton's Laws, and in particular are second-order in time; this implies momentum, and ferroelectric walls coast long distances (ca. microns) after the external electric field (E) are terminated. Second, because magnetic domain walls carry no mass, they can readily be accelerated to supersonic speeds, as shown by Democritov et al.,[1,2] at which point they emit coherent acoustic phonons at angles analogous to Cerenkov radiation or bow waves; by comparison, ferroelectric domain walls carry mass and cannot be supersonic without causing shock waves and fracture.

It has been known that ferroelectrics under HRTEM studies respond to the e-beam irradiation by significant restructuring of their domains and domain walls,[3,4] but it has not been completely clear whether this is driven thermally by beam heating and the thermal conductivity anisotropy of the target or by charging and depolarization fields. In this context it is very important to compare faceting under HRTEM with faceting observed in atomic piezo-force microscopy (PFM), since the latter does not involve the same degree of sample heating. Ganpule et al. reported a situation in lead zirconate titanate nano-structures that is probably due to thermal anisotropy along [111] axes.[5] Similar hexagonal faceting was first seen in the famous Schwartz-Hora Effect[6] and in related experiments in which laser beams produce hexagonal distributions of charged defects that fill space.[7,8] Hexagonal faceting also occurs with foams and surfactants (viscous fingering), due probably to thermal anisotropy of the substrates; but only twofold symmetry instabilities occur in magnetic bubble domains (circular to elliptical). A short pedagogic review has been given by one of the present authors,[9] and a more detailed analysis of hexagonal faceting in polyvinylidene fluoride (PVDF) films by Lukyanchuk et al.[10]

The present work seems especially interesting because the nano-crystals examined have thicknesses of ca. 80 nm which are << the length (1-3 microns) of the multi-walled nanotubes on which they are mounted, and in this respect approximate low-dimensional systems. As Berge et al. have emphasized,[11] although three-dimensional crystals are usually faceted, faceting is not permitted at thermal equilibrium in two dimensions[12] because the perimeter of a two dimensional [2D] structure is one-dimensional and cannot exhibit long-range order at finite temperatures.[13] But [2D] faceting can occur dynamically during growth processes and has been modeled numerically.[14,15] It is worth noting that unfaceted domains have been known in ferroelectrics for more than fifty years, with Cameron reporting circular "lake-like" domains in tetragonal $BaTiO_3$ in 1957.[16]

It is also important to comment on why only hexagonal faceting is observed, and not pentagons or heptagons, etc. Since the samples are single isolated films, macroscopic space-filling is not a criterion, but domain wall orientation inside the film is a criterion. Hence, there may be a relationship to the formation of foams from bubbles. Let is consider each nucleating nano-domain as analogous to a bubble, pressed nearly flat against a substrate. Only three walls meet along a line, at angles of 120° due to surface tension equality. Only four walls can meet at a point, at angles of $\cos^{-1}(-1/3) \approx 109.47°$. All these rules, known as Plateau's laws, determine how a foam is built from bubbles. Indeed, the formation of hexagonal facets in foams is well known.[17]

Our studies were carried out with a high-resolution Cs-probe corrected HRTEM (Model: JEOL JEM-2200FS) system, operated at a 200 kV voltage (~ 200 keV kinetic energy) with 0.5 A/m probe current density in order to minimize the damage rate due to Bethe-Bloch cross-section for electron-electron interaction. During the investigation, the $Pb(Zr_{0.52}Ti_{0.48})O_3$ (PZT) nanoparticles were first exposed to electron beams for an hour to get stabilized; later the images were recorded in continuous mode with the interval of 10 s. The PZT films were 50-80 nm thick, deposited on multiwalled carbon nanotubes which in turn were on n-Si substrates. The length of each nanotube was ca. 1-3 microns depending on growth conditions.[18] HRTEM studies were carried out on the PZT thin film coated multi-wall carbon nanotubes (MWCNT). Faceting behavior of PZT domains were investigated near the edge portion to maintain boundary conditions (ca. films thickness << peripheral area).

The observed faceting is global, but it is more readily seen at the edge of nanocrystals grown in the island growth common in polycrystalline PZT or $BaTiO_3$. The geometry is shown in Fig.1. In most instances the faceting was highly hexagonal with 120-degree angles between two (111) planes; however, (110) faces were also faceted. On some occasions pentagonal facets or square facets were observed. Note that the stripe domains are predominantly orthogonal to the edges in this figure. Fig.1 illustrates the domain structure at t=0 (e-beams turned on). Notice in the boxed region of Fig.1a that there is a generally round disk shape for the PZT nanocrystal, and that configurations the stripe domains inside the crystal are mostly normal to the outer hexagonal edge.

Fig.2 (a-1) illustrates the HRTEM images (both real and reciprocal space image side by side) of same target obtained in the interval of 10, 20, 30, 40, 50, and 60 seconds (s). Image taken in the next shot at t =10 s (Fig. 2(a-b)) shows the internal stripe domains that realigned predominantly parallel to the outer edge, and the hexagonal faceting is more pronounced. This suggests that the internal domain realignment controls the external faceting; inspection of transmission electron microscope (TEM) micrographs reveals that closure vertex domain structures evolve into stripe domains parallel to the external facets. In Fig.2 (c-d & e-f) shows the TEM image at t=20 & 30 s. In this condition, we see internal domain realignment and less distinct hexagonal faceting. In Fig.2 (g-h) at t=40 s shows stripe domains, in this situation (110) faces are almost orthogonal to each other. With further imaging at 50 s (fig.2 (i-j)), we observed the faceting of (110) faces. Finally imaging of same crystal at t=60 s, we observe reversion to hexagonal faceting and internal stripe domains well aligned parallel to the outer edges of the sample.

To check the universal nature of faceting, similar experiment were carried out on another target, interestingly it shows clean hexagonal faceting after continuous irradiation of e-beams. Fig. 3 shows progress of domain faceting with time. It starts with (111) parallel plane at t = 0 s, surprisingly we see 120 degree reversal of plane in next imaging time (t=10 s). Image taken in 20 s is rather more clear with evolution of (110) planes orthogonal to the hexagonal facets of (111) planes. TEM Image taken in 30 s shows the disappearance of (110) planes with clear picture of hexagonal faceting. Images obtained at 40 s and 50 s suggests further realignments of planes with interaction of energetic e-beams. The evolution of hexagonal faceting and its realignments are natural and it evolves and disappears with time, however it is not obvious that it appears and destroys with a definite interval of time.

Until the past few months domain wall oscillations in which wall velocities actually change sign were neither observed nor predicted. However, very recently Bisig et al. reported[19] changes in the sign of magnetic domain wall velocity under applied magnetic fields on a very short time-scale (100 ns) in permalloy disks of comparable geometry to the ferroelectrics in our study (50-80 nm thick; 1.0 micron radius). It is important in that work (especially their Fig.5) that the wall velocities actually reverse direction. This is interpreted as overshoot in the radial wall velocities as the domain configurations transform from vertex cores to transverse domain (stripe) walls. They record a 50 ns oscillation, about 200 million times slower than in our work. These data seem analogous to ours despite the large difference in time-scale, because we have independent evidence of both vertex (and vortex) structures in our samples[20] and of radial electric fields[3] caused by TEM charge injection. Of course the anticipated time scale for ferroelectric wall motion, involving creep velocities of typically $10^{-10}$ m/s and real mass transport, will be much slower than for spin propagation.[21]

In general, strain overshoot in materials requires viscosity and is a topic of current interest.[22] We note that in the paper by Amann et al.[22] the characteristic relaxation time for their viscoelastic materials was about 1 minute, as in the present work. We have no independent theoretical estimate of this time for our domain walls, but this comparison shows that it is similar to that in liquid crystals, which is plausible in view of the initial domain-wall topology comparison with nematics of Srolovitz and Scott. We know in the present work that the driving force for faceting is not thermal: Heating is only about 1 degree Kelvin (K).[23] The actual driving force is charging (Ahluwalia and Ng), and its effect upon surface tension. The surface tension in ferroelectric nanodomains has been analyzed by Lukyanchuk et al. [24] and shown very recently by Scott[25] to fit quantitatively hoop stress (neglected in all previous models, such as that of G. Arlt, *J. Mater. Sci.* 25, 2655 (1990)). The fact that domain wall motion in ferroelectric films can be treated as ballistic motion in a viscous medium was demonstrated clearly by Dawber et al.[26] We emphasize also that the preference for hexagon facets probably arises from the underlying lattice symmetry here with [111] axes playing a role. Although hexagonal symmetry of facets is also known for crystals with only twofold symmetry[10], strongly hexagonal faceting is observed in hexagonal magnesium (Mg) nanopores under HRTEM irradiation[23] and is registered along crystallographic axes; and Tegze et al.[27] recently report strongly hexagonal fingering in amorphous fluids (as is well known previously).

In the particular case of PZT thin-film disks, Ng et al.[28] have given a detailed model of the role of fringing fields, emphasizing that they behave quite differently for atomic force microscopy (AFM) geometries (point-like top electrode) and parallel-plate geometries, and that the 180-degree switching observed often proceeds via a two-step process involving 90-degree domains. Although the HRTEM geometry resembles AFM in the sense that there is a radial field generated by a central charge injection, the TEM beam diameter is very different from that of an AFM tip, so that different dynamics should result in these two situations. Although more detailed calculations are required, it appears that fringing fields and boundary conditions play a key role; indeed, faceting oscillations are not observed in square or triangular PZT targets.[29]

**Acknowledgement:**

Dr. Ashok Kumar (AK) would like to thank Prof. R S Katiyar, Dr. Margarita Correa & Mr. Oscar Resto for acquiring in HRTEM data.

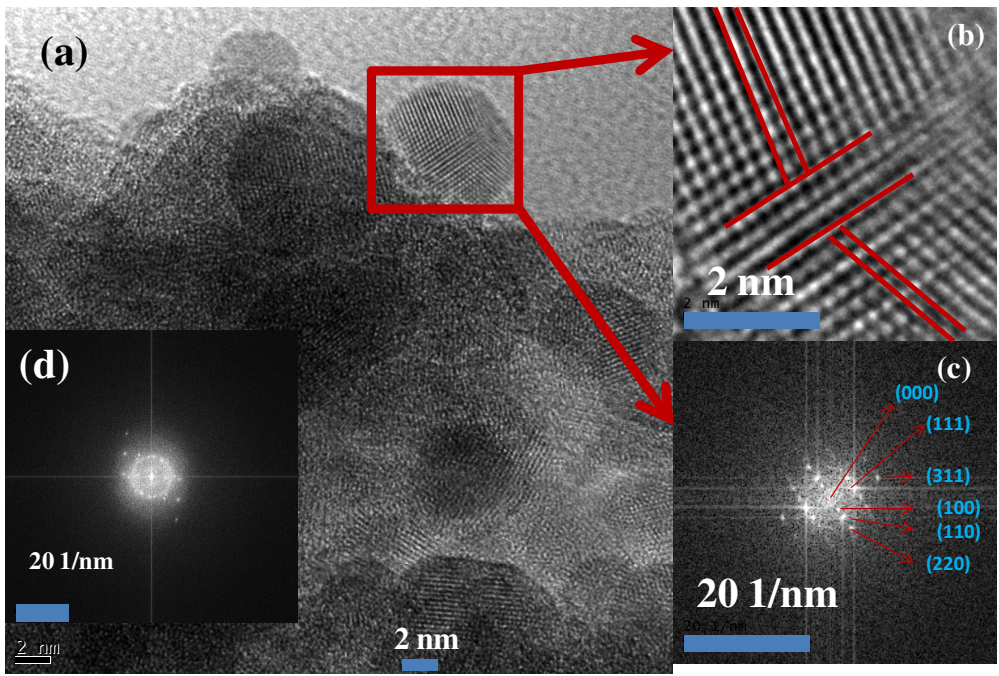

Fig. 1 Outer edge HRTEM images of thick PZT thin films (average 50-80 nm conformal coating of PZT on 1-3 micron length MWCNT): (a) Large area HRTEM image, and HRTEM image of PZT nanocrystals of sizes 5-8 nm (red box), (b) faceting of lattice plane (IFFT), (c) FFT image with assigned crystal plane, (d) FFT image of large area PZT thin films

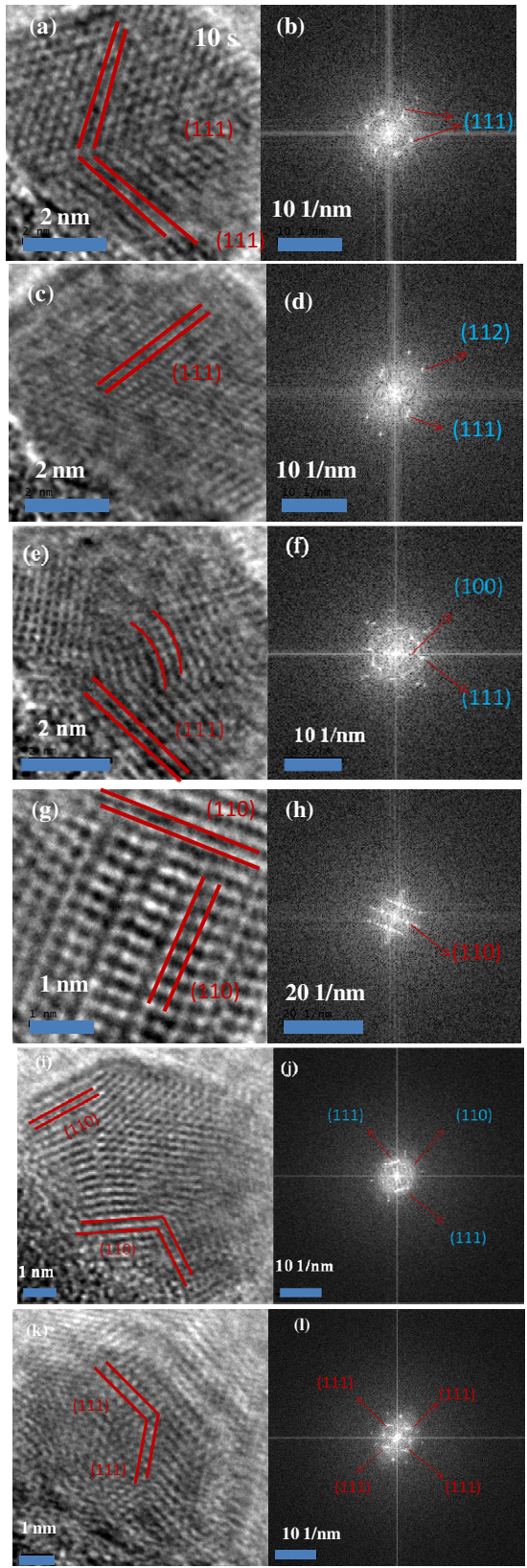

Fig. 2 (a-l) shows Inverse Fast Fourier Transform (IFFT) images and FFT images of same PZT nano-crystals (red box in fig. 1) in different time scale (10-60 s). Assigned crystal planes and their orientations are given in each figure. TEM Images were taken under the continuous irradiation of e-beams at 10 s (fig. (a-b)), 20 s (fig. (c-d)), 30 s (fig. (e-f)), 40 s (fig (g-h)), 50 s (fig. (i-j)), and 60 s (fig. (k-l)), with increasing time scale from top to bottom.

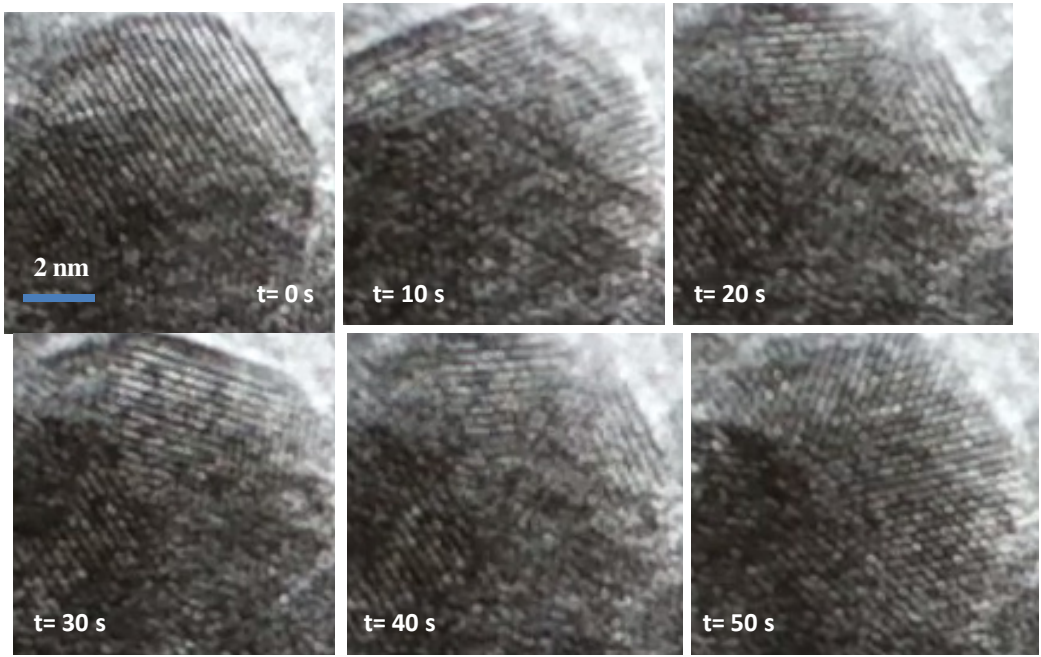

Fig. 3 shows HRTEM images of other PZT target in different time scale (0-50 s). Average size of this crystal is around 7-9 nm. Evolution and restructuring of hexagonal faceting is clearly visible with time.